\documentclass[pdflatex,sn-mathphys-num]{sn-jnl}% Math and Physical Sciences Numbered Reference Style
\usepackage{array}
\usepackage{graphicx}%
\usepackage{multirow}%
\usepackage{amsmath,amssymb,amsfonts}%
\usepackage{amsthm}%
\usepackage{mathrsfs}%
\usepackage[title]{appendix}%
\usepackage{xcolor}%
\usepackage{textcomp}%
\usepackage{manyfoot}%
\usepackage{booktabs}%
\usepackage{algorithm}%
\usepackage{algorithmicx}%
\usepackage{algpseudocode}%
\usepackage{listings}%
\usepackage{tabularx}
\usepackage{makecell}

\theoremstyle{thmstyleone}%
\theoremstyle{thmstyletwo}%

\theoremstyle{thmstylethree}%

\newcolumntype{Y}{>{\raggedright\arraybackslash}X}

\begin{document}

\title[]{The Impact of GenAI on the Future of Requirements Engineering}


\author*[1]{\fnm{Travis} \sur{Breaux}}\email{breaux@cmu.edu}

\author[1]{\fnm{Anmol} \sur{Singhal}}\email{anmolsinghal@cmu.edu}

\affil*[1]{\orgdiv{Software and Societal Systems Department}, \orgname{Carnegie Mellon University}, \orgaddress{\street{5000 Forbes Avenue}, \city{Pittsburgh}, \postcode{15213}, \state{Pennsylvania}, \country{United States}}}


\abstract{Recent advances in artificial intelligence (AI), particularly large language models (LLMs), are transforming how we design and build systems by increasing access to domain knowledge and by providing automation support to software engineering (SE). As implementation becomes less expensive through generalist SE agents, engineering effort shifts away from writing correct code and toward expressing, curating, verifying, and evaluating requirements. In this paper, we survey the state of the art in AI for requirements engineering (RE) research leading up to the transformation, before reviewing advances in LLMs. We survey two subsequent research areas: prompt programming, which treats LLM instructions as a program in SE vernacular, and generalist SE agents, which combine multiple LLM advances to yield semi-autonomous processes that complete SE tasks. Finally, we explore the future of requirements engineering along two axes: matters changing how we interact with requirements through the SE process, and matters changing how requirements are experienced by software developers and stakeholders more broadly, including end-users. This article aims to inform how RE researchers can navigate this transformation in the selection of future research priorities.}




\maketitle

\section{Introduction}

Requirements have long been a key focus of software engineering (SE). In 1970, Winston Royce argued for iterative system design because the so-called water-fall model was broken: as he described the problem, the ``testing phase'' was the first moment in software construction wherein requirements were experienced as opposed to being analyzed~\cite{Roy70}, revealing gaps in analysis that lead to discovering imperfect specifications. In the late 1980s, Barry Boehm argued for the spiral model to gatekeep development using risk analysis, prototyping, simulation, modeling, and benchmarking -- all hallmarks of rigorous requirements verification and validation practices~\cite{Boe88}. Later in the early 1990s, a new vision emerged that foresaw the automation of requirements engineering (RE) tasks through artificial intelligence (AI) built on various forms of symbolic~\cite{RW91,Ric91} and statistical reasoning~\cite{Rya93}. Yet, with the rapid and widespread adoption of the Internet in the late 1990s, it was Agile methodology that pushed back on plan-driven software development, minimizing requirements to user stories and relying on the expertise of developers to supplant the need for requirements analysis and documentation in a rapidly growing market of e-commerce and web-based systems~\cite{BT04}.

The world today is transformed by large language models (LLMs) that underpin generalist SE agents: while surprisingly unexpected, this future was foretold, but it is still arriving unevenly. This article aims to answer the question ``What is the state of the art in AI for Requirements Engineering?'' at a time when that future is rapidly evolving. In a 2025 survey of 1,006 executives, 60\% responded that, in anticipation of AI automation, they have made headcount reductions, including 21\% who have made \textit{large reductions}~\cite{DS26}. Importantly, this evidence is anticipatory: the source itself argues that layoffs are driven by AI's perceived potential rather than its demonstrated performance. Meanwhile, LLM manufacturers, including Anthropic, Google, and OpenAI, are prioritizing SE use cases with new tools to generate code. In one--possibly outdated--2026 study of 4,867 developers, the code generation assistant GitHub Copilot accounted for a 26\% increase in task completion~\cite{CDJ+26}. Some job losses represent companies shifting their capital from human developers to invest in AI infrastructure and AI cloud services~\cite{Pal26}. Although these economic developments are early industry signals described in trade press and analyst reports, they nevertheless describe a plausible future direction: a world with fewer software developers orchestrating generalist SE agents~\cite{LOG+25}. In this world, developers rely more on how well they can express requirements, curate constraints from a brownfield software development ecosystem, and evaluate agent-produced artifacts than on their ability to write correct code. Therefore, in this paper, we argue that as implementation becomes cheaper through agents, RE shifts to operational specifications, context curation, verification, and evaluation. In this regard, we see the pendulum swinging back to plan-driven development, not because risk is any greater than it was in the 1980s, but because implementation is becoming \textit{far less expensive} and automation is reducing the software engineer's \textit{visibility into code}.

To this end, we address this shifting landscape in three ways: (1) we briefly review the history of AI in RE, describing where RE researchers have primarily focused and how AI has supported that focus; (2) we review the state-of-the-art and limitations in rapidly advancing generative AI, specifically around LLMs, prompt programming, and generalist SE agents; and finally (3) we forecast how RE practice may change in an AI-driven setting where requirements increasingly instruct software agents rather than merely guiding the work of human implementers. This article is cross-cutting by reviewing how RE researchers have sought to advance the state-of-the-art in AI for RE, broadly, while situating the future of RE practice in terms of the recent disruptions caused by GenAI and LLMs, in particular. We believe that RE researchers can build on this evolving ecosystem, whether they use AI, GenAI or LLMs in any capacity in their research and methodology.

\section{History of AI for Requirements Engineering}
\label{section:history}

\begin{table}[h]
\caption{Working taxonomy of recent AI-supported requirements engineering research.}
\label{tab:summary}
\small
\setlength{\tabcolsep}{4pt}
\renewcommand{\arraystretch}{1.12}
\begin{tabularx}{\linewidth}{>{\hsize=0.5\hsize}X >{\hsize=1.5\hsize}X}
\toprule
\textbf{RE Tasks} & \textbf{Representative Works} \\
\midrule
\textbf{Elicitation} & AI-driven interviews~\cite{KGV+25, JJY+26}, script generation~\cite{TFF+26}, follow-up question recommendations~\cite{SSB25}, inquiry planning~\cite{VCG+25}, and generated feature ideas~\cite{WCL+24,WEM26}. \\
\textbf{Extraction} & Requirements extraction and classification from authoritative documents~\cite{JAS+23,AAS+22,BAF+23,DDC+23,ALC+22}, app reviews~\cite{MIG+21,SPM21,DTW+22,NKZ+23,HF23}, issue trackers~\cite{WGY+25}, and online communities~\cite{IKT+21,LSA+22}. \\
\textbf{Generation} & Domain models~\cite{SMG+21,BVD+24}, tests~\cite{AHH+24}, security requirements~\cite{KHM23,LWZ+25}, traces~\cite{DDC+23,HKC24}, goals~\cite{NCT+24,SB25}, user stories~\cite{NRK+22}, formal specifications~\cite{MMK+26}, and requirements from design guidance~\cite{DGC+25}. \\
\textbf{Analysis} & Ambiguity detection and resolution~\cite{KGH+25,HHS+21,EAA+21,EAA+22,EAA+23}, incompleteness and consistency checking~\cite{LPM22,AA24,AAB23,BKK+23}, requirements change detection~\cite{ACS+24}, violation and incompatibility detection~\cite{LZA+22}, conflict detection~\cite{FMB24}, hallucination reduction~\cite{WCJ+26}, and satisfaction checking~\cite{SBN+24}. \\
\textbf{Traceability} & Trace recovery~\cite{FHK+25}, requirements-smell analysis~\cite{VKB+25}, trace-link explanations~\cite{LLA+22}, and multi-channel requirements management~\cite{TDW+23}. \\
\textbf{Pipelines} & Operational anomaly detection from extracted requirements~\cite{AMA+22}, privacy-risk assessment from user scenarios~\cite{HKH+23}, requirements perturbation and feature tracing~\cite{GNW+23}, and test-dataset completeness evaluation from refined specifications~\cite{BRA+25}. \\
\bottomrule
\end{tabularx}
\end{table}

Requirements engineering (RE) is a broad field covering diverse concerns across a variety of systems types~\cite{Poh96}, including information-processing and cyber-physical systems. In Table~\ref{tab:summary}, RE tasks are broadly grouped into elicitation, extraction, specification, analysis, management, validation and verification, whereas AI-supported RE research has primarily covered elicitation, extraction and analysis. We distinguish between \textit{AI-assisted RE}, wherein AI is used to aid a human in the completion of an RE task, and \textit{AI-driven RE}, where the task is entirely and wholly completed by AI, even if the task was initiated by a human.

To characterize recent AI for RE research, we reviewed papers published since 2020 in prominent SE and RE journals and conference proceedings, including the IEEE Transactions on Software Engineering, ACM Transactions on Software Engineering Methodology, and the Requirements Engineering Journal, in addition to the ACM/IEEE International Conference on Software Engineering, ACM Foundations of Software Engineering, IEEE International Requirements Engineering Conference, and Automated Software Engineering Conference. This review is not systematic nor is it comprehensive. We now report our observations recent work in AI for RE.

\textbf{Elicitation.} Advances in LLMs have led to a surge of recent research in requirements elicitation. In AI-driven interviews, an evaluation of agents conducting interviews with virtual agent stakeholders found that AI makes similar mistakes as human interviewers~\cite{KGV+25}. To support human interviewers, emerging AI-assistance aims to score interview script quality~\cite{TFF+26} and generate interview follow-up questions from real-time transcript histories~\cite{SSB25}. Emerging techniques exist to hierarchically organize and prune the inquiry space to support prioritizing questions in AI-driven interviews~\cite{JJY+26} and to generate requirements from stakeholder conversations with scores for correctness, completeness and relevance~\cite{VCG+25}. Analysis comparing app-store and LLM-generated features finds that LLMs can produce useful sub-feature inspiration for novel app scopes~\cite{WCL+24}. Finally, generative image models have been used to assist app users in expressing UI improvement suggestions~\cite{WEM26}. Across this work, we observe a shift from passively capturing stakeholder statements toward active engagement via question generation and dialogue management. An open and continuing challenge is that elicitation quality still depends on surfacing tacit knowledge, that communication and human behavior underlies the effective exchange of information, and that dynamic, reactive evaluation settings are difficult to replicate.

\textbf{Requirements Extraction and Classification.} Research to extract requirements from authoritative sources includes extracting requirements and associated fragments from contracts~\cite{JAS+23}, laws~\cite{AAS+22}, requirements documents~\cite{BAF+23}, specifications~\cite{DDC+23,ALC+22}, process models~\cite{KMK+24} and GitHub issues~\cite{WGY+25}, as well as from mobile app reviews~\cite{MIG+21,SPM21,DTW+22,NKZ+23,HF23}, sub-reddits~\cite{IKT+21,LSA+22}, twitter~\cite{DTW+22} and video-based social media~\cite{SDA+23}. Related work includes emotion classification of app reviews for more enhanced feature planning~\cite{MOT+25} and user story quality assessment~\cite{MD25}. In addition, classification may include assigning requirements to the right development teams~\cite{RNP+23,BAF+23} and deciding which features to delete~\cite{NKZ+23}. Prior work also involves training small language models to improve the extraction and classification of non-functional requirements~\cite{RA25,LXX+23}. In addition, methods to promote fairness-focused RE by extracting fairness requirements from context have been proposed~\cite{FCG+24}. Three RE tasks, including classification, defect detection and conflict detection, have been reformulated as natural language inference tasks --- a format popular in natural language processing for which AI performance has improved considerably~\cite{FKH+24}. This research shows increasing breadth across artifact types and data sources, while extracted requirements exist in isolation, at best weakly connected to stakeholder intent, software project context, and downstream developer decisions.

\textbf{Generating Requirements Artifacts.} Research has used AI to generate requirements specifications and other artifacts, including traces, which can then be used for analysis. In most instances, the generated artifact is the research end and objective with the downstream task unstudied. This includes generating domain models from classroom assignments~\cite{SMG+21} and from user stories~\cite{BVD+24}, generating test cases from requirements~\cite{AHH+24}, security requirements on demand~\cite{KHM23} and from verification standards~\cite{LWZ+25}, trace links from requirements and code~\cite{DDC+23,HKC24}, context-sensitive user stories from diverse specifications~\cite{NRK+22}, safety goals from a malfunction catalog and scenarios~\cite{NCT+24}, goal models from interview transcripts~\cite{SB25}, formal specifications from natural-language requirements~\cite{MMK+26}, and requirements from design guidelines and designer question-answering~\cite{DGC+25}. Unlike the extractive research, wherein extracted artifacts maintain strong lexical and semantic coherence with the source data, generative research relies on synthesis and translation to yield greater interpretation and conversion between concepts present in the source data and those emerging in the target artifacts. For example, while user stories provide evidence of user actions and goals, scenarios that one could generate from such stories can include additional assumptions (e.g., about the kinds of data and data inter-dependencies that must exist to support those actions), which are missing or not explicitly stated in the original stories.

\textbf{Disambiguation, Inconsistency and Completeness.} Requirements analysis leads to improving requirements and specifications by identifying, removing, and augmenting defects in the form of ambiguities, inconsistencies and omissions. Research has used AI to identify ambiguity in feature requests~\cite{KGH+25}, financial specifications~\cite{VSB23} and system policies~\cite{HHS+21}, identify missing links in issue trackers~\cite{LPM22}, changes in requirements found in laws~\cite{ACS+24} and to detect requirements violations and incompatibilities~\cite{LZA+22}. AI has been used to detect and interpret anaphoric ambiguities~\cite{EAA+22}. To resolve ambiguity, external, supplemental domain knowledge can be introduced~\cite{EAA+21,EAA+23}. Neuro-symbolic methods showcase how to combine language models with SMT-based satisfiability reasoning to detect conflicting requirements~\cite{FMB24}. Where ambiguity and completeness refer to what is missing, hallucinations in language model usage refer to what is inaccurate and inconsistent. Use-case, entity-relationship and create, read, update, and delete (CRUD) models can be used to reduce hallucinations in LLM-based requirement auto-completion~\cite{WCJ+26}. More recently, researchers have started to assess the downstream impact of clarifying ambiguous requirements on code generation~\cite{MSW+24,TC25}. In addition, research has used AI to translate natural language requirements into models for consistency and completeness checking~\cite{AAB23,BKK+23}, to check completeness of data processing agreements~\cite{AA24} and to verify whether natural language specifications satisfy requirements~\cite{SBN+24}.

\textbf{Traceability and Requirements Management.} A significant body of work focuses on automatically recovering trace links among requirements, code, documentation and user feedback. This includes using retrieval-augmented generation for generic traceability link recovery and improves code-related traceability tasks~\cite{FHK+25}. Requirements smells have a measurable effect on LLM-based traceability decisions, suggesting that prompt quality and requirements quality can interact in automated trace recovery~\cite{VKB+25}. Other work generates and visualizes trace-link explanations to help non-experts understand why artifacts are related~\cite{LLA+22}. A requirements ecosystem study links forum posts, issue tracker entries and FAQs, showing how deep learning can support requirements management across user-feedback channels ~\cite{TDW+23}. Traceability is continuing to provide mechanism for retrieval and explanation across iterative development of heterogeneous artifacts, while challenges persist in how to establish and maintain trusted links, particularly as software engineering becomes increasingly generative.

\textbf{Integrated Analysis Pipelines.} While most RE research reviewed above has studied narrowly-scoped RE tasks, a few projects have sought to create pipelines spanning multiple RE task categories. For example, aviation requirements were extracted from forum posts and product documentation and used to detect anomalies in real-time flight operations~\cite{AMA+22}. In other work, users author scenarios about their use of software, which were next analyzed by AI to identify privacy sensitive data types that users then scored for privacy risk~\cite{HKH+23}. Existing requirements were perturbed to identify novel variations, which were then traced to existing features and studied through software development team reflections~\cite{GNW+23}. Operational design domain attributes, such as road and lighting conditions for autonomous vehicle testing, were refined into specifications by AI, which are then used to evaluate test dataset completeness~\cite{BRA+25}. Whereas narrowly scoped tasks are amenable to ground-truth evaluation with a single dataset, pipeline integration must contend with the reality of cascading errors, in which imperfect artifacts or data identified by AI are used as input to a downstream task, which is then evaluated separately.

\section{Advances in Language Models}

Since the introduction of the third generation Generative Pre-Trained Transformer (GPT-3), followed by the public inflection point around ChatGPT and GPT-3.5, large language models (LLMs) have undergone significant advances impacting nearly every industry in modern civilization. Many LLMs are instruction-tuned and aligned, which consists of fine-tuning and reinforcement learning to follow human instructions~\cite{WBZ+22} and maximize helpfulness, honesty and harmlessness~\cite{ABC+21}. Early work on FLAN-style instruction tuning found substantial improvements in zero-shot task performance~\cite{LHV+23}. Few-shot chain-of-thought (CoT) prompting raises awareness that models can improve on reasoning tasks when examples include intermediate reasoning steps~\cite{WWS+22}; zero-shot CoT, in which users prompt an LLM to ``Let's think step by step,'' was found to improve performance accuracy on math problem tasks by 30-50\%~\cite{KGR+22}.

Today, frontier language models (Claude, Gemini and GPT) include ``native'' multimodal models trained on text, images and code, and large reasoning models (LRMs) trained to generate long chain-of-thought. Although the exact sizes and training costs of leading closed models are not generally disclosed, frontier models are widely understood to require large-scale training runs involving many GPUs and substantial capital expenditure. Through model distillation, in which a large model is used to train a small model, a variety of ``small'' language models averaging in size from 1B to 12B parameters exist, which can be further fine-tuned on single-GPU workstations. Finally, modern models can make use of a \textit{mixture of experts} (MoE), in which model layers are partitioned into multiple ``experts'' and a routing gate selects a subset of experts for each input~\cite{JJN+91,JJ94}. This allows the experts to specialize through training on smaller parts of a problem, which reduces training and inference costs since only a subset of the entire network is active at any one time.

\begin{figure}[htbp]
    \centering
    \includegraphics[width=0.8\textwidth]{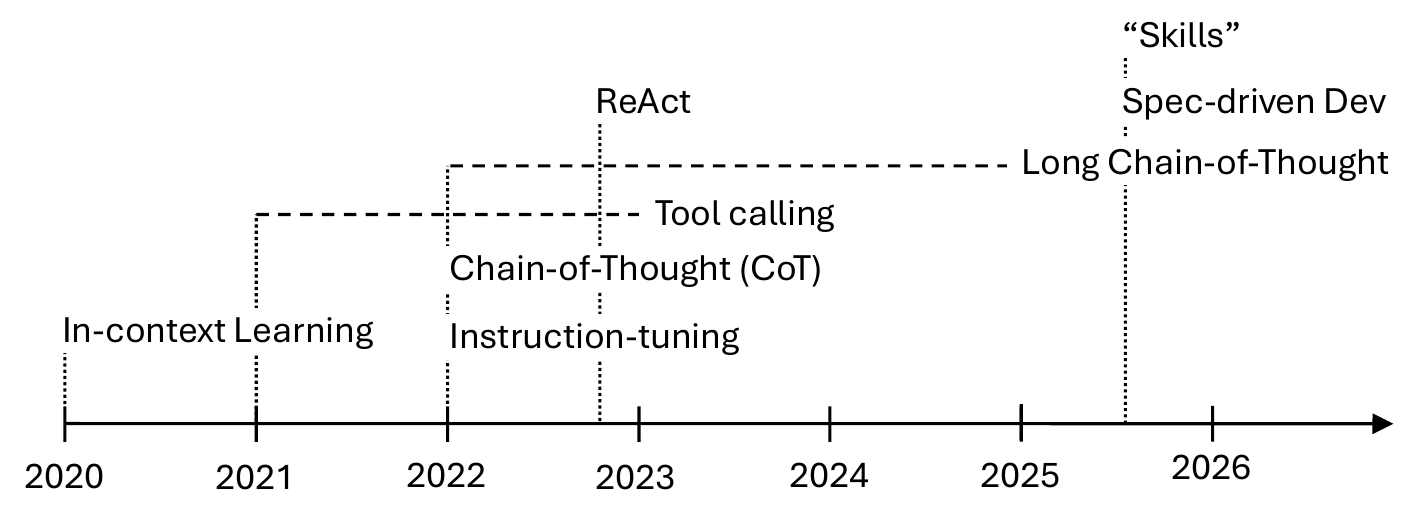}
    \caption{Timeline of significant advances in large language models.}
    \label{fig:llm-timeline}
\end{figure}

Several observations emerge over the state-of-the-art that affect how recent advances in AI and LLMs in particular can impact RE research and practice. In Figure~\ref{fig:llm-timeline}, we present a general overview of several significant advances in LLM research and development that we discuss further in this section. Following the introduction of GPT-3 in 2020~\cite{BMR+20}, chain-of-thought (CoT) prompting emerged in early 2022~\cite{WWS+22} and instruction-tuning was introduced with InstructGPT in the same year~\cite{OJA+22}. Tool-calling was first observed in the 2021 WebGPT study~\cite{NHB+21}, after which it was later embedded in LLM APIs in 2023 through GPT-3.5 and GPT-4; today, tool calling is standard post-training for agent-supporting LLMs. Long CoT, which was inspired by CoT and introduced through a post-training method called STaR in 2022~\cite{ZWM+22}, first arrived in production in late 2024 with GPT-o1. The agentic loop, introduced by the popular ReAct framework~\cite{YZY+23}, appeared in late 2023~\cite{YZY+23}. Most recently, ``skills'' that describe re-usable plans appeared in mid 2026 alongside Amazon Kiro and GitHub SpecKit. Taken together, these advances enabled the creation of software engineering (SE) agents and spec-driven development, which we discuss at the end of this section.

\subsection{Pre-training Limitations} 

The quality and type of pre-training data determines which ``skills'' can be learned later~\cite{ZLX+23}. GPT-3 was a notable advance not just because of its size (175B parameters), but because its pre-training data included highly curated prose from web crawl data, WebText2 pages linked from Reddit posts, and vast quantities of fiction and non-fiction describing human emotion, motivation, empathy and strategy. This early data also had its limitations: numerical representations of multiplication and division were largely absent and there was an absence of code, which corresponded to poor math performance~\cite{BMR+20} and the inability to generate executable code~\cite{CTJ+21}. Finally, due to pre-training \textit{cut-off dates} -- the date after which the LLM parameters are frozen and no new information is learned -- there will always be a need for augmented generation~\cite{GXG+23}, whether using retrieval-augmented generation~\cite{FDN+24} or tool calling~\cite{PZW+24,QLY+24,YSR+24}. In fact, the under-representation of knowledge in pre-training is directly proportional to inaccuracy in zero-shot question-answering~\cite{MAZ+23}.

In RE, tacit knowledge, also referred to as the \textit{unknown knowns}~\cite{SGN11}, refers to knowledge that a stakeholder has, but which they withhold from the development process for reasons including that they do not realize they have it or that they struggle to articulate it~\cite{GGR+13}. Requirements elicitation interviews are the predominant technique for surfacing tacit knowledge~\cite{FSG+16}. A consequence of tacit knowledge is under-representation of such knowledge in modern pre-training data: unspoken or unwritten stakeholder needs are unlikely to be captured in web text, audio or video. Consequently, because the accuracy of LLM-based question-answering is proportional to the number of relevant pre-training documents~\cite{MAZ+23}, accuracy on tasks concerning tacit knowledge is likely to be poor. Moreover, long-tail knowledge, which is largely under-represented~\cite{KDR+23}, leads to increased incidence of hallucination~\cite{ZGM+25}, which means LLMs can generate factual inaccuracies or inconsistencies when LLMs are prompted on topics covering tacit knowledge. While many measures of language models have converged on fluency~\cite{HBC+20}, we also know that human trust increases as reading effort decreases, independent of whether the information is valid~\cite{RS99,LK10}, which can lead to over-reliance and misplaced trust in generative AI~\cite{GW20}.

\subsection{Symbolic and Statistical Reasoning} 

Language models struggle to emulate logical reasoning~\cite{CLL+25}, which is the foundation of many RE analysis methods~\cite{LMS+24}. For example, when LLMs are tested on symmetric relations learned during fine-tuning, e.g., testing ``B is A'' when ``A is B'' was learned, LLM performance is no better than random chance~\cite{BTK+24}. However, translating natural language problems into logic has since improved by decomposing and chaining translations of atomic formulae needed to construct complex logic~\cite{RKL+25}, and then fine-tuning LLMs on synthetic logic samples to improve accuracy by 9-11\% on both logic and code benchmarks~\cite{MMY+24}. In contrast, chain-of-thought prompting~\cite{WWS+22} raises awareness that inference-time compute to generate reasoning trajectories can increase accuracy in math word problems. This observation led to frontier-scale large reasoning models (LRMs) that generate long chain-of-thought (LCoT), which is encapsulated at inference time and later truncated from the response (e.g., in a generated $<$think$>$$</$think$>$ block). While LCoT simulates cognitive behaviors (e.g., verification, backtracking and sub-goal setting) in reasoning~\cite{GCS+25}, the improvements gained by CoT have been limited to logic and algorithmic reasoning tasks, with performance on classification and question-answering tasks no different from random chance~\cite{SYR+25}. In LCoT, the cost of increasing inference-time compute is prohibitive, particularly when LRMs are known for overthinking, which can arrive at and later skip a correct answer in the reasoning trajectory~\cite{ZMC+25}. Finally, it's important to note that CoT is not an explanation, nor is it necessarily faithful to the generated conclusion~\cite{TMP+23} and unfaithfulness increases as language models become sufficiently large~\cite{LCR+23,BSM24}.

Reasoning by requirements analysts has traditionally been inquiry-focused~\cite{PTA94} to guide requirements elicitation, and model-driven notations and formalisms, including use cases, data flow diagrams, state-transition diagrams, goal models and business process models~\cite{Poh96}. While research using formal methods have studied the benefits of automation to assist the human analyst~\cite{tCC+24}, and logic is part of a historical foundation of AI reasoning~\cite{GN12}, there remains a gap between the recent advances in LLM-based reasoning and reasoning used to perform requirements validation and verification. Neuro-symbolic reasoning is a promising advance that combines the statistical, translation capabilities of LLMs with the formal guarantees of theorem proving and model checking~\cite{FDZ+24}. This can be used to translate natural language requirements into formal specifications expressed in logic, called auto-formalization~\cite{WJL+22}, which can then be checked using model checkers and verifiers~\cite{WFC+25}.

\subsection{Context Management} 

LLMs have evolved to include ever increasing context windows from approximately 2K tokens in GPT-3 to 8K-32K tokens in GPT-4-era models and 128K-1M tokens in some modern frontier models. Larger contexts enable new tasks that require processing large documents and code bases without pre-filtering this data. Context augmentation, which consists of adding relevant data to the context window for task completion, is a strategy to introduce new information that emerged after the LLM cut-off date and to reduce hallucinations. This includes both retrieval-augmented generation~\cite{GXG+23} and custom retrievers~\cite{FDN+24} and tool calling~\cite{QLY+24}, in which LLMs generate tool calls to collect information through APIs~\cite{PZW+24}. However, large contexts increase inaccuracy in task performance when the information needed to answer questions becomes ``lost in the middle''~\cite{LLH+24}, including a leading bias in summarization tasks~\cite{RSC+24}. Attention loss is particularly a problem in long-horizon tasks (LHTs), which consist of extended sequences of steps with delayed feedback and frequently combine context augmentation with reasoning to complete multi-step problem solving. 

\section{Requirements Re-visited}

Large language models (LLMs) have led to two new ways that researchers think about requirements. In prompt programming, language model ``prompts'' are viewed as programs that include ``requirements'' intended to constrain the model output. In software engineering (SE), there has been a shift toward deploying a version of a product, however imperfect, and relying on bug reports and feature requests to allow the product to better meet stakeholder needs post-deployment. This process shift highlights the value of LLM-based agents in the resolution of project issues. Finally, to guide SE agents in general software development tasks, specification (spec)-driven development has emerged as a means to anchor agents on documented requirements, which serve as input to the agent. We now discuss these approaches as we re-visit emerging ways that software engineering engages with requirements.

\subsection{Prompt Programming}
Prompt programming is a new paradigm that emerges from human computer interaction and treats the LLM input context as a natural language program~\cite{FKC+23,RM21} that introduces a new class of requirement, defined as ``a skeleton instruction that communicates an essential condition or constraint on desired LLMs output''~\cite{MPY+25}. In prompt programming, ``requirements engineering'' is the practice of iterating over the instructions to improve the output, for example, by ignoring implicit requirements, such as ``respond in the same language,'' that need not be stated, and by making task-relevant requirements explicit, such as ``correct the grammar in this e-mail''~\cite{MPY+25}. Programmers report frequently having to identify and introduce unknown knowns or tacit knowledge into prompts to correct unintended LLM output~\cite{LLR+25}. Debugging is an exercise in revising the prompt program to localize faults~\cite{LYS+25}. In prompt programming, quality requirements refer to the prompt output and may be induced by role instructions, e.g., ``software developer who writes clean and simple code''~\cite{KOM+25}. Where prompt programming tasks concern code generation, the conventional definition of requirement overlaps this new class, e.g., when a prompt instruction defines a requirement for the behavior of the generated code~\cite{LLR+25}.

\subsection{Advances in Agentic AI}
Agentic AI integrates recent advances in LLMs and LRMs into a \textit{harness} that consists of sensors, actuators and control loops. 

\subsubsection{Generalist Agents}
The thought-action-observe (TAO) loop introduced by ReAct~\cite{YZY+23} is the foundation of modern LLM-based agents, in which CoT is generated to plan the next action, after which the result of the action is observed for feedback. If feedback includes observed errors, the next thought step can choose an alternative action and re-try~\cite{SCG+23}. Sensors and actuators correspond to the inputs and outputs of tool calls, wherein a tool is an API function, which can further access command-line programs and scripts. Tools can be selected during decoding~\cite{SDD+23} from a list of API references~\cite{QLY+24} or agents can build their own tools through code generation~\cite{XWY+25}. Because LLMs are highly sensitive to context changes~\cite{SCT+24}, small changes to tool descriptions can impact tool selection reliability~\cite{FWC+25}. Inspired by robotics, LLM-based agent performance improves when planned actions are grounded in the world~\cite{ABB+22,HXX+22}, which has been extended to solving computer tasks, including using command-line and web-based interfaces~\cite{KBM23}. Skills compose tools into text-based procedures with ``explicit applicability conditions, execution policies, termination criteria, and reusable interfaces,''~\cite{JLD+26} which support agents in solving longer horizon tasks~\cite{CGZ+26}.

Multi-agent systems build on a simple paradigm, which consists of \textit{proposers} that generate content, and \textit{aggregators} that synthesize, summarize, and discriminate signals from generated content~\cite{WWA+25}. Agents can assume roles~\cite{LHI+23}, which specialize their generation and synthesis tasks, particularly during agent debate~\cite{DLT+24,CCS+24}.

\subsubsection{Long-Horizon Tasks}
Long-horizon tasks (LHTs) can lead to compounding errors as incorrect output from one step propagates as input to the next step in the sequence. Agent \textit{gyms} and \textit{arenas} have emerged to test an agent's abilities in LHTs and discover limitations, including in web page navigation~\cite{LGL+25,ZXZ+24}, desktop operating system usage~\cite{XZC+24}, GitHub issue resolution~\cite{JYW+24} and programming tasks~\cite{JGL+25}. Unlike classical benchmarks, gyms and arenas operate ``live'' environments consisting of a containerized operating system configured with running programs. Agents can be assigned a ``budget'' to limit token utilization and other sources of under-performance, such as task re-try attempts. Evaluation metrics include task completion~\cite{LYZ+24} and tool selection~\cite{QLY+24,PZW+24} accuracy, and agent-to-agent matchup performance via win rates and ELO ratings~\cite{CZS+24,ZCS+23}. Because task trajectories are non-deterministic, pass@k is a popular metric, which estimates the probability of at least one successful outcome across \textit{k} attempts~\cite{CTJ+21}. There is an increasing over-reliance on LLM-as-a-Judge~\cite{ZCS+23} to perform evaluations of two outcomes, wherein little established ground truth presupposes that LLMs are sound and reliable verifiers, even if they are flawed generators. This over-reliance is in contrast to recent evaluations that show LLM-as-a-Judge with strong models produce biased labels~\cite{DNH25} and perform slightly better than random guessing~\cite{TZM+25}.

\subsubsection{Software Engineering Agents}
Software engineering agents tailor the generalist agent paradigm to software engineering tasks, such as issue resolution and debugging. For example, issue resolving agents use tools to find and edit relevant files in project repositories before submitting a pull request containing a code patch that resolves a reported issue~\cite{WLS+25,YJW+24}. Popular benchmark are curated from GitHub project repositories to include Python code bases, issues and pull requests~\cite{JYW+24,ZHZ+26} and executable runtime environments~\cite{PWN+25}, as well as, a pipeline to generate training data for benchmarks~\cite{YLJ+25}. Evidence shows that hand-crafted AI pipelines for issue resolution can still outperform synthetic data and Agentic frameworks~\cite{XDD+25}.

\subsection{Spec-Driven Development}

Spec-Driven Development (SDD) re-frames requirements specifications as operational inputs to coding-generating agents in contrast to being reference documents only reviewed by humans developers~\cite{Pis26}. In SDD, specifications or ``specs'' are a version-controlled, behavior-oriented artifact that describes product scenarios, user stories, EARS-formatted requirements~\cite{MWH+09}, environmental and development constraints, and expected outcomes before implementation begins~\cite{DR26}. Specs, which in RE sit at the intersection of the machine and the environment~\cite{Jac95}, more broadly cover requirements artifacts in SDD and are intended to provide the coding agent with the necessary context to steer implementation. The SDD defines a spectrum in which specs gain increasing authority over code: in \textit{spec-first} development, the spec is written before coding to guide the initial implementation, which may drift away from the spec over time; in \textit{spec-anchored} development, the spec is maintained alongside the code and alignment is enforced through testing; and in \textit{spec-as-source} development, developers edit specifications while code is generated from the spec~\cite{Pis26}.

Recent tools impose SDD workflows on development, including GitHub SpecKit~\cite{DR26}, Amazon Kiro, BMAD, and OpenSpec \footnote{\href{https://kiro.dev/}{Amazon Kiro}, \href{https://github.com/bmad-code-org/bmad-method}{BMAD} and \href{https://github.com/Fission-AI/OpenSpec/}{OpenSpec}}. Each workflow defines a spec format, level of detail, and how specs are integrated into a project. While spec files are written in natural language, early research argues that LHTs require additional artifacts, including scenarios, models, and Gherkin-style specifications to improve SDD outcomes~\cite{FCM+26}. 

The SDD vision and current tool support have open challenges in software development process quality, including a lack of sufficient evaluation frameworks and benchmarks to verify requirements satisfaction and traceability. LLMs struggle to generate executable behavioral specifications, especially at repository level, where the best model in CodeSpecBench solves only 20.2\% of tasks~\cite{CDZ+26}.  Maintaining spec satisfiability over time is also challenging because, as the software project grows, the code could drift away from the specs on which the code was initially based. Studies of repository-level context files show mixed results: while agents.md files, which contain agent instructions, rules, and context, may reduce median runtime and output token use while preserving completion behavior~\cite{LMG+26}, when context files contain unnecessary requirements they are observed to have the opposite effect, reducing task success and increasing token use~\cite{GMM+26}. Similarly, explicit plans can improve issue resolution by agents, whereas subpar plans and poorly aligned instruction language can degrade performance~\cite{LDG+26}. As prior work in prompt engineering has shown, LLMs are sensitive to input contexts and research is needed as task complexity increases. 

%Therefore, a potential research direction for AI in RE is to translate elicited requirements into minimal, structured, and prioritized specifications: separating mandatory constraints from preferences, ordering work by dependency and risk, and maintaining traceability from each prioritized requirement to plans, generated code, and tests.

\section{Future of Requirements}

Advances in AI create new opportunities for the future of RE research. In particular, we focus on two broad dimensions: matters of software process, including how AI is changing the way that software is constructed; and matters of experiencing requirements, which shifts how we collect, analyze, and realize requirements in software.

\subsection{Matters of Software Process}
In the early aughts at the turn of the century, Agile methodology fundamentally changed how requirements were integrated into software development by reducing effort invested into requirements planning and emphasizing the need to work with skilled developers~\cite{BT04}. Arguably, the growing prevalence of frameworks and reusable code enabled this change by accelerating development when developers accept requirements from third-party code. SE agents have the potential to again transform this integration by shifting attention back to well-defined requirements, because agents increasingly sit between the developer and the code. While Agile methods depend heavily on expert developers to function without extensive documentation and planning~\cite{BT04}, SE agents require upfront planning and direction, including instructions with detailed requirements~\cite{Pis26}. In this section, we review the ways that LLMs are transforming the requirements engineering process.

\subsubsection{From Validation to Verification}
Recent AI-assisted RE tasks surveyed in Section~\ref{section:history} primarily cover \textit{requirements validation}, which concerns whether organizations have the right requirements by acquiring requirements-related data from novel sources~\cite{IKT+21,AAS+22,LSA+22,DTW+22,SDA+23} and identifying requirements defects~\cite{HHS+21,LPM22,AAB23,BKK+23,KGH+25}. Less work covers \textit{requirements verification}, which concerns checking whether design specifications and code satisfy requirements~\cite{AMA+22,SBN+24}, as well as work to bridge requirements to verification, including test case generation~\cite{AHH+24} and trace link discovery~\cite{DDC+23,HKC24}. Meanwhile, developers routinely face challenges using AI-assisted code completion, including difficulty interpreting and debugging generated code~\cite{VZG22,BJP23}, which obfuscates the information needed to perform verification. Developers report that requirements are often too abstract to include in prompts, instead they have to first decompose requirements into programming tasks manually, which are then further enriched by developers with design decisions and architectural constraints~\cite{UKV25}. 

Requirements analysis and implementation planning by future SE agents will require intentional fine-tuning of LLMs to be proficient, as evidenced by NLP performance gains realized through instruction-tuning~\cite{LHV+23}. While zero-shot chain-of-thought prompting was an emergent LLM property~\cite{KGR+22}, long chain-of-thought, which underpins LRMs and includes inferential behaviors such as verification and backtracking~\cite{GCS+25}, was only achieved through reinforcement learning~\cite{ZWM+22}. SE agents introduce additional abstractions between the developer and code, introducing new questions about the maintainability and reliability of specs in agentic development. New work is needed to evaluate SE agents in requirements comprehension and to extend SE agentic benchmarks to cover requirements task evaluation. 

Spec-driven development proposes to create a path from validation to verification by turning validated requirements into executable or checkable artifacts. If a valid spec is used to shape and control generated artifacts in downstream tasks, then generated code, generated tests, and generated documentation must be checked against behavior intended by the spec. Preliminary research already shows that executable behavioral specifications encode pre-conditions and post-conditions that can accept valid behavior and reject invalid behavior~\cite{CDZ+26}. Auto-formalization may further increase access to formal methods by translating natural language mathematics into formal specifications and proofs~\cite{WJL+22}, while neuro-symbolic methods may allow offloading parts of verification to theorem provers and model checkers. 

\subsubsection{Singular to Long-Horizon Tasks}

\label{section:downstream}
Prior work has largely focused on well-defined, separable, and singular RE tasks, for example, deciding which software features to delete~\cite{NKZ+23}. However, the efficacy of task automation is determined by introducing the task output into a downstream task, such as identifying code-level dependencies on features targeted for removal, accepting or rejecting the feature removal request based on these dependencies, removing code associated with accepted feature deletion requests, and updating requirements documentation to reflect the change. In prior research, the downstream task is frequently presented as motivation using vague terminology, for example, to support requirements elicitation and analysis~\cite{KMK+24}, to support maintenance and version tracking~\cite{WGY+25}, or to support product improvement, customer outreach, and competitive analysis~\cite{MIG+21}. A consequence of this vagueness is that the downstream task is under-specified and unclear, in which case the singleton task output may not be fit-for-purpose.

RE therefore needs long-horizon benchmarks that evaluate requirements utility beyond singular tasks. SE has SWE-bench-style issue-resolution benchmarks for assessing whether agents can modify real repositories~\cite{JYW+24}. However, RE does not have a widely accepted benchmark suite where agents must elicit and analyze real requirements, plan an implementation against those requirements, maintain traces, and test requirements satisfiability. To leverage recent advances in agentic AI, future RE research should focus on RE-specific arena and gym designs to include: \textit{elicitation tasks}, such as detecting tacit knowledge or design trade-offs that require clarification of stakeholder intent; \textit{modeling and analysis tasks} to surface implied requirements or requirements conflicts; and \textit{trace preservation tasks} to avoid losing requirements and design intent over multiple cycles of agentic loops and tying implementation back to stakeholder acceptance tests. With LHT evaluation, arenas and gyms move beyond testing only whether a requirements artifact is fluent. In practice, we understand that the best benchmarks are: (1) based on real data, not synthetic data; (2) designed with a deterministic, measurable objective that is used to show improvement; and (3) so difficult that baseline attempts achieve roughly 10\% accuracy, which is small enough to leave room for improvement and large enough to show that machine learning is feasible.

Advances in LLM-based code generation introduce new opportunities to extend RE tasks into implementation and testing. In testing, for example, a long-horizon task consists of identifying requirements, finding relevant code segments distributed across multiple files that relate to those requirements, writing and executing test cases, and changing the code to pass those test cases. Code can represent traditional requirements abstractions, including models~\cite{JHG+24}, which may support translating the results of model-based analysis into code generation and evaluating the effects of this translation through novel, long-horizon task benchmarks. Original requirements analysis techniques that pre-date the latest AI advances, such as model checking~\cite{CAB+98}, model merging~\cite{NSC+07}, and obstacle analysis~\cite{VL00,AKV+12}, may be re-imagined as RE tools for use by SE agents. A challenge, however, is managing cascading error across the duration of LHTs and supporting tool calls, which requires innovation in evaluation.

To take advantage of these LHT opportunities, RE researchers must invest in acquiring more data for LLM training and fine-tuning, particularly in rare and specialized domains. Underlying every agent harness used to solve an LHT is an LLM that has been pre-trained on the domain and the sub-tasks required to fulfill the LHT. The absence of requirements documentation in public code repositories elevates reverse engineering research to acquire requirements from code~\cite{YWM+05}. While some might argue that the gold standard is industry data in RE, the code generating model Codex was fine-tuned from a GPT-3 base on public GitHub repositories~\cite{CTJ+21}, which are frequently criticized for being mostly inactive or personal projects~\cite{KGB+16}. This is because, in addition to memorization, LLMs learn patterns and clone behavior from data, which can generalize to solve industrial tasks that fit the same patterns as public, open source projects. In contrast, recursive training on synthetic data is known to lead to model collapse~\cite{SSZ+24}. This observation is in contrast to model distillation, which is shown to yield high performance in small models trained from data generated by larger models~\cite{GCS+25}.

\subsubsection{Innovation in Evaluation}

As instruction-tuned LLMs are increasingly used to manipulate and transform the requirement text, and as requirements are further integrated into downstream tasks, novel metrics and benchmarking methods are needed to evaluate the effectiveness of future AI-driven RE. The traditional natural language generation (NLG) metrics BLEU and ROUGE, for example, have known limitations that reduce their effectiveness in evaluating generated requirements artifacts. BLEU measures modified n-gram precision with a brevity penalty~\cite{PRW+02}, and is known to measure improvement without increases in quality assessed by human judges~\cite{COK06,Rei18}. ROUGE is a family of recall-oriented overlap metrics~\cite{Lin04}, and can exhibit measured improvement when the generated text contains hallucinations~\cite{BGM+19,MNB+20}. In code generation, BLEU and ROUGE penalize programs with structural dissimilarity that produce the same, correct output~\cite{CTJ+21}. LLM-as-a-Judge~\cite{ZCS+23} also does not solve this problem. Agentic benchmarks are increasingly relying on LLM judges, but recent evaluations show that LLM judges can produce biased labels~\cite{DNH25} and perform only slightly better than random guessing in certain settings~\cite{TZM+25}. 

Human evaluation is also insufficient when it is dominated by fluency. NLG researchers documented over 200 metrics used across 478 single-criteria evaluations in a survey of 165 research papers from 2000-2020~\cite{HBC+20}. Among these metrics, fluency (``how grammatical and readable is the text?’’) mapped to the largest number of normalized criteria, which is known to correlate with human trust~\cite{RS99,LK10}: as fluency increases in the text, humans are more likely to trust the text, independent of what the human may need to comprehend from the text. Moreover, human evaluators applying an evaluation rubric to score factuality tend to employ a fluency heuristic by rating fluent text as higher quality regardless of factual content~\cite{CAS+21,JHN+23}. Wherein LLMs are used to solve increasingly complex, long-horizon tasks, researchers risk subjecting their evaluations of generated artifacts to fluency bias, reducing their ability to measure the construct of interest. In RE, for example, fluent requirements and fluent trace explanations may appear plausible to human evaluators, even when they fail to preserve stakeholder intent or satisfy a specification.

Due to the dominance of fluency as a mediating factor, requirements evaluation may need to shift from human assessments of generated requirements quality (e.g., ambiguity, completeness, consistency) to measuring \textit{requirements utility}, or how useful and fit-for-purpose generated requirements artifacts are in LHTs. Agent \textit{gyms} and \textit{arenas} introduce LHT evaluation strategies in web page navigation~\cite{LGL+25,ZXZ+24}, desktop operating system usage~\cite{XZC+24}, GitHub issue resolution~\cite{JYW+24} and programming tasks~\cite{JGL+25} that measure changes in outcomes. For example, GitHub issue resolution benchmarks~\cite{JYW+24} use pass@k~\cite{CTJ+21}, which estimates whether at least one of \textit{k} agent-generated code samples passes oracle test cases, while the effects of agents browsing the web and completing web forms are evaluated using content metrics for \textit{exact match}, \textit{string subset}, and \textit{fuzzy match}~\cite{ZXZ+24}. Moving beyond requirements text and quality metrics requires a deeper understanding of how changes in requirements affect changes in the state of software development.

\subsubsection{Requirements Evolution \& Management}
Software engineering (SE) agents and Spec-Driven Development (SDD) have the potential to change requirements management from a repository of stored requirements into an active process for evolving agent-facing artifacts. When specifications guide implementation, a change request should first be interpreted as a proposed change to existing spec files. The scenarios, constraints, plans, tests, and code affected by the changes can then be updated as downstream artifacts. This is especially important in brownfield projects, where the agent must preserve existing behavior while adding or revising functionality~\cite{Pis26}.

Requirements management is not only to store specifications, but to keep specs minimal, current, and performant in SE agent frameworks. Context files (e.g., agents.md and spec.md) reduce development time or increase task failure, depending on how they are written~\cite{LMG+26,GMM+26}, particularly if they contain irrelevant information~\cite{GMM+26}. To address evolution, context files must be versioned and maintained, with frequent changes by adding or modifying instructions, sections, and testing guidance~\cite{MGT+26}. To understand and predict the impact of requirements evolution on SDD workflows, we need new methods for change-impact analysis over spec-related artifacts. The goal is to detect conflicts early, propagate accepted changes to the correct artifacts, and preserve traceability across repeated AI-driven software revisions.

A key challenge going forward will be context management by ensuring that SE agents that are refactoring code adhere to previously documented requirements, assumptions, and constraints. As this documentation grows, new techniques for managing and scoping only relevant context to a given task will be needed to avoid attention loss. Vanilla retrieval-augmented generation (RAG) methods may be insufficient during a LHT, instead requiring recursive summarization~\cite{WFC+25}, hierarchical memory management~\cite{SZZ26}, or graph-based retrieval techniques~\cite{LZH+25}, among others, adapted to scope and trace requirements to partitions of software project repositories.

\subsection{Matters of Experiencing Requirements}

Recent advances in AI have the potential to shift how requirements are experienced by developers and stakeholders, including users. As upfront requirements elicitation, documentation and analysis have been reduced over time in Agile development, developers have come to experience stakeholder ``needs'' through software implementation: as they author code, they realize needs for process concurrency to improve response times, or needs to pre-process and stage data to simplify user interfaces and workflows~\cite{EM12}. Such needs arise during coding due to under-specified requirements, which developers have managed to backfill while they face implementation challenges or observe performance deficits during implementation and testing. This experiential process illustrates Jackson's argument $S, K\vDash R$, wherein requirements are a by-product of software and knowledge, regardless of whether the requirements were collected and documented early in the process. 

We organize this section in three parts: developer experience of requirements during implementation will be reduced as they off-load code authorship to AI; with AI, developers have easier access to preview requirements through behavioral artifacts, such as prototypes, that have traditionally been too costly to implement; and stakeholders could have new opportunities to express requirements through more responsive and adaptive AI-driven means.

\subsubsection{Reductions in Authorship Experience}
Advances in AI are changing how requirements are experienced by developers: developers who rely on SE agents to complete SE tasks will defer their experiences of requirements later into testing, in memory of Royce's critique of the waterfall model~\cite{Roy70}. The traditional practices of interpreting requirements to plan implementation, of authoring code to implement the plan, and of writing test cases to verify that the authored code will satisfy requirements are being delegated to SE agents (e.g., Claude Code or GitHub Copilot). Historically, as developers author software components during implementation, they often experience unstated assumptions in requirements. For example, integrating a component with an audio processing library can introduce audio sampling and encoding constraints imposed by the library, which may have been unspecified in the original requirements; or segmenting data processing into asynchronous components can raise unstated performance requirements when components run at different speeds and slower than anticipated. In Agile software development, these unknown knowns may be identified and resolved in the code by the developer, while remaining undocumented. With SE agents, developers will off-load these experiences to agents, which will cause their comprehension of requirements to become shallow,  less particular and less technical. Developers need greater transparency into requirements-related design choices that arise during implementation so that those choices may be reviewed, supervised, and documented. Meanwhile, written requirements and specifications could emerge as a contract between the developer and agent, if the effects of that documentation leads to guarantees of software behavior in development and production. 

\subsubsection{Previewing Behavioral Experiences}
An important consequence of LLMs today and going forward is the reduced effort needed to author benchmarks, prototypes, and simulations in code. Benchmarks provide a means to evaluate and compare alternative design choices against established performance measures and to test for performance regression, while prototypes allow developers to reduce risk by exploring technology integrations. In software engineering research and practice, SE agents can now be used to generate multiple, divergent prototypes for far less cost and time, allowing stakeholders to review and select the best prototype. After learning about the effectiveness of integrations to satisfy requirements, prototypes can be thrown away, if they represent a path to abandon, or preserved as part of the system to-be~\cite{Dav02}. Simulations are used to test scalability and to evaluate the effects of edge cases, particularly in autonomous systems~\cite{BKR+25}. Combined, these techniques allow developers and other stakeholders to preview, assess and experience various requirements outcomes faster, even before the requirements are explored or understood. Advances in LLMs to generate code can be used to expedite development and re-introduce these planning activities originally proposed in plan-driven development~\cite{Boe88}. These techniques may be used to define requirements-based guardrails around development spaces: the outcome of design variances introduced by SE agents, for example, may be observed through benchmarks, explored using prototypes, and assessed at scale using simulations.

\subsubsection{Responsive and Adaptive Specification}
As Agile methods have minimized costly planning procedures, techniques have emerged to preserve and simulate access to stakeholder needs, often through stakeholder surrogacy. This includes the product owner role, which is a member of the development team who represents stakeholder values and intent~\cite{SB01,BBR+18}, and personas, which are representative ``cutouts'' used to motivate and guide discussions about how stakeholders view and experience needs~\cite{PA10}. The extent to which SE agents become increasingly responsible for larger portions of the software development life cycle (SDLC), the cost of customization and personalization may be reduced to allow end-users to more easily communicate their unmet needs directly to designers, SE agents or both. In RE, traditional ML has been used to mine user needs from app reviews~\cite{LL17,DLP+22}, which capture delayed descriptions of user need, often long after an unmet need is experienced. In contrast, SE agents create the possibility that users can identify their unmet needs in the moment of experience, wherein the salience of those needs is highest and the description of desired outcomes may be more easily demonstrated and evaluated by those users as SE agents iterate over solutions. This includes capturing \textit{ephemeral requirements} that require satisfaction within minutes or hours of experience~\cite{FBM02,LPN12}. To accommodate real-time refactoring, software must be designed to support modifiability, evolution, and even self-adaptivity, in which software sensors dynamically detect the need to undergo design adaptation~\cite{CLG+09,DGM+13,GWQ21}.

\section{Conclusion}

Requirements engineering (RE) activities have long been considered repetitive and tedious due to the intensive human effort required to collect, analyze, and synthesize under-specified and conflicting needs into actionable requirements. In the past five years, a transformation through generative artificial intelligence (AI) has impacted nearly every domain in which software is used. Against the backdrop of an outstanding rate of innovation in software engineering (SE), we observe a shift in the use of AI from solving singular tasks to longer-horizon tasks, in which SE agents situate code generation into agentic planning loops using observation, feedback, and verification. This innovation creates new opportunities for re-introducing historically heavy-weight RE methods into the software engineering pipeline, while allowing developers to see the outcome of RE-related changes throughout the pipeline. Future RE tasks must account for the underlying technology fundamentals and innovation in large language models and agentic AI. While the future is hard to predict in this rapidly changing moment, we envision two general ways that requirements stand to change: in how requirements are integrated into software processes, shifting the focus from validation, which remains relevant, toward verification, especially as a form of evaluating automation over long horizon tasks; and in how requirements are experienced, shifting developer attention away from code authorship and toward evaluating design alternatives and observing the behavioral effects of selected designs, while increasing stakeholder access to directly evolve requirements and implementations at their own time of need.

\end{document}